# Wavelet-based Heat Kernel Derivatives: Towards Informative Localized Shape Analysis

M. Kirgo[1,2], S. Melzi[1,4], G. Patanè[3], E. Rodolà[4] and M. Ovsjanikov[1]

[1]LIX - École Polytechnique - IP Paris, [2]EDF R&D, [3]IMATI CNR, [4]University of Rome La Sapienza

**Abstract**
*In this paper, we propose a new construction for the Mexican hat wavelets on shapes with applications to partial shape matching. Our approach takes its main inspiration from the well-established methodology of diffusion wavelets. This novel construction allows us to rapidly compute a multiscale family of Mexican hat wavelet functions, by approximating the derivative of the heat kernel. We demonstrate that this leads to a family of functions that inherit many attractive properties of the heat kernel (e.g., local support, ability to recover isometries from a single point, efficient computation). Due to its natural ability to encode high-frequency details on a shape, the proposed method reconstructs and transfers δ-functions more accurately than the Laplace-Beltrami eigenfunction basis and other related bases. Finally, we apply our method to the challenging problems of partial and large-scale shape matching. An extensive comparison to the state-of-the-art shows that it is comparable in performance, while both simpler and much faster than competing approaches.*

## 1. Introduction

In the last decade, advances in 3D shape analysis have seen the emergence of a class of methods falling under the umbrella of *diffusion geometry*. Based on the seminal work of Coifman and Lafon [CL06], such approaches leverage the relation between the geometry of the underlying space and the diffusion process defined on it, as encoded especially by the spectrum of the Laplace-Beltrami operator (LBO, for short). This general strategy has been successfully exploited for the construction of point signatures [SOG09, GBAL09] and shape matching [OBCS*12] among other tasks. More recently, progress in this field has shifted towards a more "local" notion of shape analysis [OLCO13, MRCB18], where descriptors are computed only on small and properly selected neighborhoods (Sect. 2). This choice is motivated by several relevant settings dealing with real-world 3D data, where the acquired shapes have missing subparts, due to self-occlusions, or a wildly different mesh connectivity. To date, however, combining informative diffusion-based geometric techniques with robust localized shape analysis has remained an elusive goal addressed by few methods [OMMG10, HVG11, HQ12, OLCO13, MRCB18].

In this paper, we propose an extension to the classical diffusion-based constructions by considering functions that are obtained as time derivatives of the heat kernel (Sect. 3). Such functions have local support, thus providing a natural tool for capturing multi-scale shape properties. Furthermore, they inherit fundamental properties of the heat kernel [OMMG10], such as an efficient computation together with the ability to recover isometries from a single point. Our construction is also related to Mexican hat wavelets that we build directly on the surface while avoiding spectral approximations.

From a functional standpoint, the resulting family of functions forms an over-complete basis (a *frame* or, as we refer to below, a *dictionary*) that provides a richer functional representation power, compared to standard LBO eigenfunctions or heat kernel functions. For example, delta functions supported at surface points are reconstructed more faithfully through our representation under a lower memory budget (Sect. 5). This aspect has direct consequences in several applications, such as dense correspondence, function transfer across shapes, and partial shape matching (Fig. 1 and Sect. 6).

Our contributions can be summarized (Sect. 7) as follows:

- we introduce heat kernel derivatives as a novel tool for localized shape analysis;
- our proposed representation is compact and efficient to compute, while allowing an accurate representation of Mexican hat wavelets. Furthermore, it is demonstrably competitive with full-fledged algorithmic pipelines for partial shape correspondence and similarity, at a fraction of the computational cost;
- we compare our approach to popular Mexican hat wavelet formulations and prove that it achieves the best trade-off between efficiency and accuracy.

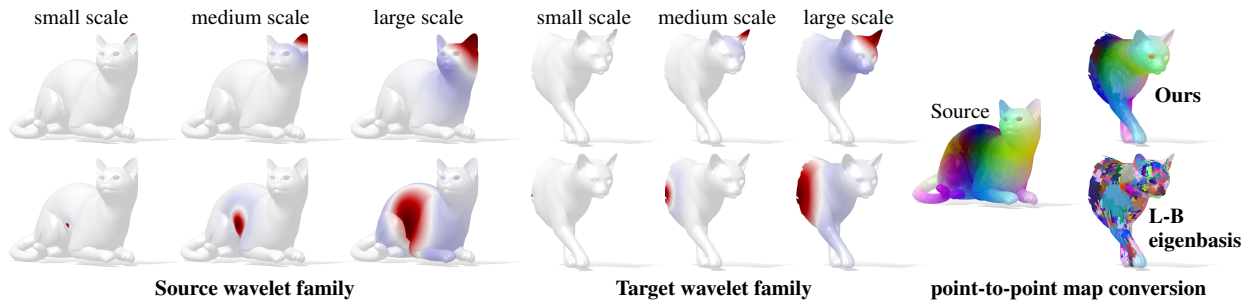

**Figure 1:** *Example functions from the proposed wavelet family on a pair of shapes, respectively a full model (left) and a partial near-isometry (center). Each function is represented at three scales (from left to right) and localized around two different samples (1 for each row). The rightmost column shows the point-to-point correspondence obtained using our wavelet family (top) and the standard Laplace-Beltrami (L-B) eigenbasis (bottom). Both maps are estimated on the same set of 13 landmarks, and visualized by color coding. Our construction is designed to be the preferred choice in the partial setting.*

## 2. Related work

The definition of a compact and efficient representation of signals is a fundamental task in geometry processing. By far, the most common approach is to use the eigenfunctions of the Laplace-Beltrami operator, which are a natural extension of the Fourier basis to surfaces [Tau95, Lév06, VL08]. In most settings, a truncated approximation consisting of the low frequency eigenfunctions is used to guarantee numerical robustness and computational efficiency. The LBO eigenfunctions basis lies at the core of many global and pointwise shape signatures, such as [Rus07, RWP06, SOG09, GBAL09, ASC11, MRCB16] and has been widely used for shape deformation [RCG08], segmentation [RBG*09], and functional transfer [NMR*18]. In [ABK15], Aflalo *et al.* showed that the LBO eigenfunctions are the optimal for representing continuous functions with bounded variation, thus providing a theoretical justification for its versatility.

The LBO eigenfunctions are also commonly used in the functional map framework [OBCS*12], which relies on approximating and transferring functions in reduced bases. Despite its prevalence, the truncated LBO eigenfunction basis suffers from three main limitations: (i) the support of its functions is global [MRCB18, OLCO13], (ii) the truncated set of eigenfunctions provides a low-pass filter on the signal and thus it is not able to approximate functions composed of high frequencies [NMR*18], and (iii) the LBO eigenfunctions are defined up to sign and suffer from switches in the sign and the order, even for near isometric shapes [SK14].

To address these challenges, several alternatives to the LBO eigenfunction basis have been proposed. In [KBB*13], Kovnatsky *et al.* define a compatible basis on shape collections by performing a simultaneous diagonalization of the LBO. The compressed manifold modes [NVT*14, OLCO13, KGB16] provide a set of sparse and localised basis functions. Modifying the LBO, Choukroun *et al.* [CSBK17] define a Hamiltonian operator, whose eigenfunctions are localized in those regions that correspond to the modification of the LBO. In [MRCB18], a similar solution is applied to define a basis that is also orthogonal to a given set of functions.

In [NMR*18], Nogneng *et al.* use polynomial combinations of the LBO eigenfunctions basis in conjunction with standard linear combinations of functions to allow the representation of higher frequencies. In [MMM*20], the LBO eigenfunctions are extended using the coordinates of the 3D embedding. The resulting "Coordinates Manifold Harmonics", capture both extrinsic and intrinsic information, encoded in the standard LBO basis. Our use of an overcomplete functional dictionary is also related to the recent Binary Sparse Frame [Mel19], where a set of non-orthogonal indicator functions improve the approximation and the transfer of step functions through sparse coding. Finally, a set of diffusion and harmonic bases have been proposed recently in [Pat18], based on properties of the heat kernel.

**Local and multi-scale processing** More closely related to our work are multi-scale shape analysis methods [HPPLG11] with (i) local descriptors [Joh99, CJ97, BMP01, HSKK01, MHYS04] and (ii) diffusion geometry [SOG09, VBCG10, BK10, Pat13, PS13, Pat16]. These latter methods typically exploit the multi-scale nature of the heat kernel, which captures progressively larger neighborhoods of a given point while being able to characterize local geometry efficiently. However, the signatures based on heat diffusion can fail to capture important (e.g., medium frequency) shape details, which has led to other descriptors, such as the Wave Kernel Signature [ASC11] and optimal spectral descriptors [Bro11, WVR*14].

While these approaches focus on the *discriminative power* of the computed descriptors, wavelet-based techniques aim explicitly to construct locality-aware functional families. With respect to the spectral graph wavelet signature [LH13] and the spectral graph wavelet transform [HVG11], our approach does not rely on an eigen-decomposition and solves a small set of sparse linear systems, which allows to capture local details and to operate on complex geometries. We provide an extensive comparison with the most closely related wavelet methods in [HVG11] (Sect. 3.2) and show that our approach leads to a rich functional family that can be computed more efficiently compared to [HVG11], while capturing local high frequency details, crucial for partial shape matching.

**Wavelets on surfaces** Finally, our work is inspired by the construction of wavelet-based functional families on triangle meshes [Zho12, Ch. 4] based on subdivision [LDW97], diffusion [CM06], and eigendecomposition [HVG11]. While our work does not fit directly in this field, we base our construction on diffusion wavelets and specifically propose to consider the negative time derivative of the heat kernel to construct our multiscale functional family. In the Euclidean domain, this time derivative (or equivalently second derivative in space) corresponds to the Mexican hat wavelet. Moreover, since it is constructed without relying on the LBO eigendecomposition [HQ12], it provides a very efficient and powerful tool for local shape analysis.

## 2.1. Background & motivation

Our main goal is to construct a family of functions that is both *local* and provides a *multi-scale description of the shape geometry*, analogously to wavelets in Euclidean domains. The most classical approach for generating a family of wavelet functions is *via* shifting and dilation (or scaling) of a generating function, referred to as the *mother wavelet*. Extending this approach to curved surfaces is challenging because shifting and dilation are not canonically defined on non-Euclidean domains. As a result, a large number of approaches [CM06, HQ12] circumvent these challenges by replacing these operations with those easier to mimic on surfaces.

Our construction is based on the notion of *diffusion wavelets*, which broadly exploit the link between diffusion and function dilation. As a way of motivation, consider a standard zero-mean Gaussian function on the real line: $f_0(x) = (\sigma\sqrt{2\pi})^{-1} \exp(-x^2/(2\sigma^2))$. If $f_0$ is dilated and re-scaled by $1/s$, then we obtain another Gaussian $\frac{1}{s} f_0(x/s)$, whose standard deviation is multiplied by $s$. On the other hand, if we consider a diffusion process $\partial_t f(x,t) = \partial_{xx}^2 f(x,t)$, then its fundamental solution is given by the classical heat kernel $f(x,t) = (4t\pi)^{1/2} \exp(-x^2/(4t))$. Recalling that the heat kernel satisfies $f(x, \sigma^2/2) = f_0$ and noting that $f(x, s^2\sigma^2/2)$ is a Gaussian with standard deviation $s\sigma$, we get that: $f(x, s^2\sigma^2/2) = \frac{1}{s} f_0(x/s)$. This computation shows that in certain cases, dilation and scaling can be equivalently computed by *solving the diffusion equation* starting with $f_0$. While the above computation is done with a Gaussian function $f_0$, a similar result also holds for the Mexican hat (Ricker) wavelet, which is defined as the negative second order derivative of a Gaussian function.

According to these observations, the key idea of diffusion wavelets [CM06, HQ12] is to *replace dilation by diffusion*, which is particularly useful on curved surfaces. In fact, while defining dilation is itself difficult, diffusion is well defined by replacing the Laplacian $\partial_{xx}^2 f(x,t)$ with the Laplace-Beltrami operator. Following this line of work, our main goal is to construct a multi-scale family of functions that both have strong locality properties and exhibit good approximation of other functions through linear combinations. Previous approaches have exploited these links either by using a multi-scale family built directly from the heat kernel [CM06] or by operating in the spectral domain, through a truncated representation [HQ12]. Instead, we build a multi-scale functional family using the derivative (in time or, equivalently, in space) of the heat kernel and operate purely in the spatial domain by explicitly simulating heat diffusion. This choice allows us to both avoid an expensive eigen-decomposition necessary to approximate very local functions and to achieve better function reconstruction accuracy, exploiting the multi-resolution properties of the Mexican hat wavelet.

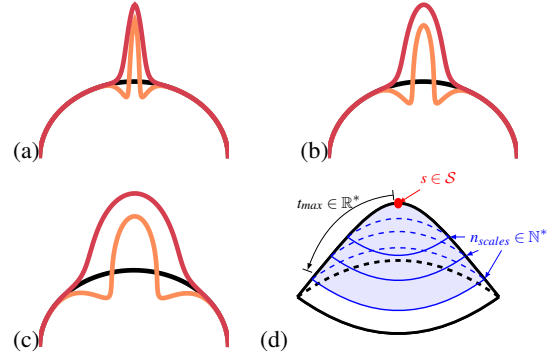

**Figure 2:** *(a-c) Illustration of our diffusion wavelets (orange) on a 1D manifold (in black), and corresponding "scaling functions" (red), which approximate the heat kernel evaluated at the same sample. (d) Parameters of our approach: largest diffusion time $t_{max}$, number of scales $n_{scales}$, and samples $s$. The end of the support of successive wavelets $\left\{\psi_{s,n}^M\right\}_{n \in [1; n_{scales}]}$ is represented by blue lines and the light blue region is covered by the wavelet at $t_{max}$.*

## 2.2. Continuous setting

In the spirit of [HQ12], we define a wavelet-like family by means of a diffusion process on a manifold $\mathcal{M}$ (Fig. 2(a-c), red curve). The resulting family provides a dictionary of functions and the corresponding linear vector space spanned by them is used for the representation of a function in wavelet coefficients. Let $u : \mathcal{M} \times \mathbb{R} \to \mathbb{R}$ be the solution to the heat equation

$$\partial_t u(x,t) = -\Delta u(x,t), \quad u(x,0) = u_0(x), \quad t \in \mathbb{R}^+. \quad (1)$$

If the initial condition is defined at a single point (i.e. $u_0(x) = \delta_y(x)$ with $y \in \mathcal{M}$), then the solution of Eq. (1) is the heat kernel $\mathbf{K_t}(x,y)$. $\mathbf{K_t}$ provides a family of Gaussian-like functions on the surface $\mathcal{M}$, with increasing standard deviation (or increasing "scale") as $t$ grows. At a given scale, the negative first-order derivative of such a function constitutes the diffusion Mexican hat wavelet. Equivalently, the Mexican hat wavelets $\psi_t(x,y)$ at scale $t$ can be computed from the heat kernel $\mathbf{K_t}(x,y)$ as follows:

$$\psi_t(x,y) = -\partial_t \mathbf{K_t}(x,y) = \Delta_x \mathbf{K_t}(x,y), \quad (2)$$

where $\Delta_x$ denotes the Laplace-Beltrami operator with respect to the point $x$.

Given the Laplacian eigensystem $\{\lambda_k, \Phi_k\}_{k=0}^{+\infty}$, an exact spectral formulation of the heat kernel in the continuous setting exists and is given by:

$$\mathbf{K_t}(x,y) = \sum_{k=0}^{\infty} \exp(-t\lambda_k) \Phi_k(x) \Phi_k(y). \quad (3)$$

Therefore, the associated family of wavelets is defined as:

$$\psi_t(x,y) = \sum_{k=0}^{\infty} \lambda_k \exp(-t\lambda_k) \Phi_k(x) \Phi_k(y). \quad (4)$$

In [HQ12], this property is used to define Mexican hat wavelets

**Algorithm 1** Computation of a dictionary of Mexican hat-like functions for a set of samples $\mathcal{S}$. $\Omega_{\mathcal{M}}$ is the area of $\mathcal{M}$. $A_{\mathcal{M}}$ and $W_{\mathcal{M}}$ designate the normalized area and cotangent weight matrices, computed using Neumann boundary conditions. $\rho \in (0;1]$ is an adjustment ratio, and $||.||_1$ the $L_1$ norm w.r.t. $A_{\mathcal{M}}$.

**Input:** set of samples $\mathcal{S}$, number of scales $n_{scales}$, maximal diffusion time $t_{max}$, ratio $\rho$
**Output:** $\Psi_{\mathcal{S}}$ (multi-scale dictionary for all $s \in \mathcal{S}$)
  $t \leftarrow \rho \frac{t_{max}}{n_{scales}\sqrt{\Omega_{\mathcal{M}}}}$
  $\Psi_{\mathcal{S}} \leftarrow \{\}; \quad \psi_{\mathcal{S}}^{\mathcal{M}} \leftarrow A_{\mathcal{M}}^{\dagger} W_{\mathcal{M}} \delta_{\mathcal{S}}; \quad \psi_{\mathcal{S},0}^{\mathcal{M}} \leftarrow \psi_{\mathcal{S}}^{\mathcal{M}}$
  **for** $n \leftarrow 1$ to $n_{scales}$ **do**
    $\psi_{\mathcal{S},n}^{\mathcal{M}} \leftarrow (A_{\mathcal{M}} + tW_{\mathcal{M}})^{\dagger} A_{\mathcal{M}} \psi_{\mathcal{S},n-1}^{\mathcal{M}}; \Psi_{\mathcal{S}} \leftarrow \{\Psi_{\mathcal{S}}, \psi_{\mathcal{S},n}^{\mathcal{M}}\}$
  **end for**
  Normalize each column $c$ of $\Psi_s$ with $||c||_1$
  Normalize each column $c$ of $\Psi_s$ by $max(c) - min(c)$

on $\mathcal{M}$ as a truncated version up to $N = 300$ LBO eigenpairs:

$$\psi_t(x,y) = \sum_{k=0}^{N} \lambda_k \exp(-t\lambda_k) \Phi_k(x) \Phi_k(y), \quad (5)$$

In this work, we construct a dictionary of localized functions, based on the same intuition but avoiding the eigen-decomposition, and instead solving the diffusion equation directly. Our dictionary shares the following properties (Fig. 2d) with the spectral Mexican hat diffusion wavelets:

- it is based on the same defining relation (through derivatives in time or space) between the heat kernel and the Mexican hat wavelets as in the Euclidean setting;
- our functions are located at a set of chosen sample positions $\mathcal{S}$, and the resulting dictionary provides a multi-scale representation via a maximum diffusion time $t_{max}$ and a chosen number of scales $n_{scales}$.

## 3. Proposed approach

We introduce our construction of diffusion wavelets on discrete surfaces (Sect. 3.1). We compare their accuracy to other diffusion wavelet constructions (Sect. 3.2), their conversion to a point-to-point map (Sect. 3.3), and analyze their main properties (Sect. 3.4). In Sect. 4, we perform an in-depth empirical study of these properties.

We assume that shapes are represented as triangle meshes in the discrete setting. We also assume that each shape $\mathcal{M}$ is endowed with the Laplace-Beltrami Operator $L_{\mathcal{M}} = A_{\mathcal{M}}^{\dagger} W_{\mathcal{M}}$, where $A_{\mathcal{M}}$ and $W_{\mathcal{M}}$ are respectively the area and cotangent weight matrices [MDSB03] and $A_{\mathcal{M}}^{\dagger}$ is the pseudo-inverse of $A_{\mathcal{M}}$.

### 3.1. Discrete setting

For convenience, in the following we consider the case with one sample location at vertex $s$. To build a dictionary of functions at various scales, we make three observations.

1. Given the $\delta$-function at $s$, denoted $\delta_s$, in the discrete setting, one can compute a Mexican hat wavelet by applying the LBO to $\delta_s$.

2. Given a function $f$, one can compute a scaled version of $f$ by applying the diffusion operator $D_t$ to $f$. Additionally, the "scaling factor" is controlled by the diffusion time $t$.

3. $D_t$ can be approximated precisely and efficiently via a backward-Euler scheme.

Observation 1. follows from the relation between the heat kernel and the Mexican hat wavelet summarized in Eq. (2) and the fact that the Laplace-Beltrami and diffusion operators commute. In other words, computing the heat kernel $\mathbf{K_t}(s,x)$ and then applying the LBO to obtain $\psi_s(x) = \Delta \mathbf{K_t}(s,x)$ is equivalent to computing $D_t \Delta \delta_s$. This approach leads to an analogue of the "mother wavelet" and provides the means to save computational effort, since it avoids applying the Laplacian to each scale of the heat kernel independently. In practice, the mother wavelet is obtained by computing $\psi_s^{\mathcal{M}} = A_{\mathcal{M}}^{\dagger} W_{\mathcal{M}} \delta_s$, where $A_{\mathcal{M}}$ and $W_{\mathcal{M}}$ are computed using Neumann boundary conditions and the vertex coordinates of $\mathcal{M}$ are divided by $\Omega_{\mathcal{M}}$ which is the total area of $\mathcal{M}$. By applying Observation 2, for increasing diffusion times $t$ to $\psi_s^{\mathcal{M}}$, we obtain a set of multi-scale Mexican hat-like functions $\Psi_s = \left\{\psi_{s,n}^{\mathcal{M}}\right\}_{n \in [1;n_{scales}]}$. Finally, Observation 3 provides an efficient way to compute 2. In practice, we found that $10 - 50$ Euler-steps allow to approximate $D_t \psi_s^{\mathcal{M}}$ better than a truncated spectral formulation (Sect. 3.2).

Moreover, we directly use each intermediate function $\psi_{s,n}^{\mathcal{M}}$ obtained at the $n$-th backward Euler step as a function of our dictionary. In other words, the number of scales $n_{scales}$ represents the number of backward-Euler steps that we use to produce the Mexican hat wavelet associated to a diffusion time $t_{max}$. Given a function $f$ on $\mathcal{M}$, applying one backward-Euler step to approximate the effect of the discretized diffusion operator $D_t$ amounts to computing the quantity $f_{diff} = (A_{\mathcal{M}} + tW_{\mathcal{M}})^{\dagger} A_{\mathcal{M}} f$. Therefore, given the $(n-1)$-st wavelet at a sample $s$, we compute the $n$-th wavelet as : $\psi_{s,n}^{\mathcal{M}} = (A_{\mathcal{M}} + tW_{\mathcal{M}})^{\dagger} A_{\mathcal{M}} \psi_{s,n-1}^{\mathcal{M}}$.

In our applications, we use a linear time sampling: $t = \rho \frac{t_{max}}{n_{scales}\sqrt{\Omega_{\mathcal{M}}}}$. If two shapes $\mathcal{N}$ and $\mathcal{M}$ are involved in the computation, the ratio parameter $\rho$ is set to $\rho = \frac{\sqrt{\Omega_{\mathcal{N}}}}{\sqrt{\Omega_{\mathcal{M}}}}$. $\rho$ adjusts the diffusion scales on the two shapes so that they relate well in practice. If a single shape is involved, $\rho = 1$. This ratio is especially useful in the case of partial shape matching, where $\mathcal{N}$ is the partial shape and $\mathcal{M}$ the full shape.

Storing the set of sample locations $\mathcal{S} = \left\{s_1, ..., s_{|\mathcal{S}|}\right\}$ in the matrix $\delta_{\mathcal{S}}$ (the $k^{\text{th}}$ column of this matrix is $\delta_{s_k}$) allows us to compute the wavelets at all sample locations in parallel: instead of computing a single mother wavelet $\psi_s^{\mathcal{M}}$, we compute a set of mother wavelets, stored as the columns of a matrix $\psi_{\mathcal{S}}^{\mathcal{M}}: \psi_{\mathcal{S}}^{\mathcal{M}} \leftarrow A_{\mathcal{M}}^{\dagger} W_{\mathcal{M}} \delta_{\mathcal{S}}$, which are propagated *via* backward Euler steps to $t_{max}$. This procedure (Algorithm 1) enables us to compute the full dictionary $\Psi_{\mathcal{S}}^{\mathcal{M}} = \Psi_{s_1}, ..., \Psi_{s_{|\mathcal{S}|}}$ efficiently.

### 3.2. Comparison to other wavelet formulations

Our approach has two main competitors: the Mexican hat wavelets [HQ12] and the spectral graph wavelets [HVG11]. We also compare our construction to an alternative approach using the wFEM diffusion operator [Pat13], which replaces the backward

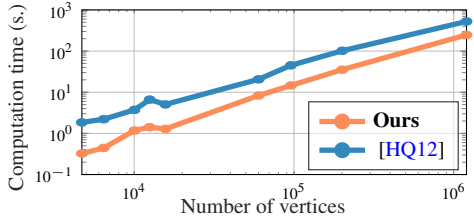

**Figure 3:** *Comparison of the scalability of eigen-decomposition-based wavelet computation method [HQ12] and our approach. The wavelets are computed at 10 sample locations, using 25 scales.*

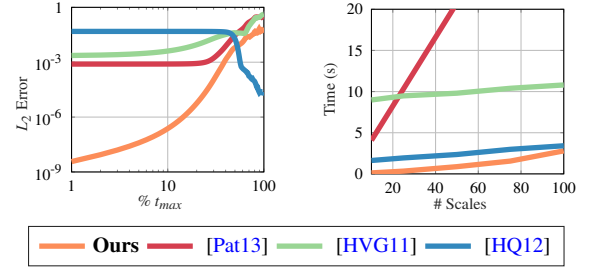

**Figure 4:** *Comparison between various Mexican hat diffusion wavelet definitions and our approach as $L_2$ error to ground truth wavelets (left) and computation time (right), on the complete set of 100 shapes of the FAUST data set (remeshed to shapes with 5K vertices). See the averaged values in Table 1.*

**Table 1:** *Comparison of four Mexican hat wavelet formulations on the FAUST data set (100 remeshed shapes with 5K vertices). The $L_2$ and $L_\infty$ norms are used as accuracy error measures.*

|  | **Ours** | [Pat13] | [HQ12] | [HVG11] |
| --- | --- | --- | --- | --- |
| Av. $L_2$ | $\mathbf{1.7 \times 10^{-2}}$ | $9.7 \times 10^{-2}$ | $2.3 \times 10^{-2}$ | $9.9 \times 10^{-2}$ |
| Av. $L_\infty$ | $\mathbf{9.7 \times 10^{-2}}$ | $6.3 \times 10^{-1}$ | $5.1 \times 10^{-1}$ | $7.1 \times 10^{-1}$ |
| Av. t.(s) | **1.14** | $2.2 \times 10^1$ | 2.46 | 9.89 |

Euler approximation scheme in the generation of the Mexican hat wavelets. Note that we do not compare our method with the work of Coifman *et al.* [CM06], which defines orthogonal wavelets, but not Mexican hat wavelets, using a diffusion operator. Moreover, their construction does not allow to select a set of samples from which to compute the wavelet functions, whereas we rely on this information. Finally, their method is performed *via* a full bottom-up approach, starting from all vertices of the considered shape, to large-scale orthogonal wavelet functions. This process uses a costly iterative procedure that is not well suited to our applications that involve dense meshes.

To illustrate the scalability of our approach compared to methods leveraging an eigen-decomposition of the Laplace-Beltrami operator, we measure in Fig. 3 the time required to compute a dictionary at 10 sample locations and 25 scales for [HQ12] and our method. We use 8 shapes from the SHREC'19 data set (connectivity track) [MMR*19], – see the Appendix (Sect. 8.2) for details on the data sets used in our experiments – with an increasing number of vertices, to which we add an additional shape of around 1.2M. vertices, produced by applying the Catmull-Clark subdivision method [CC78] to the largest shape of this data set. A table of the computed values is provided in the Appendix (Table 7, Sect. 8).

Our comparison to the competing definitions is based on three criteria: (i) $L_2$ error to the ground truth Mexican hat wavelets (Fig. 4 (left), Table 1), (ii) $L_\infty$ error to the ground truth Mexican hat wavelets (Fig. 14 of the Appendix, Table 1), (iii) computation time (Fig. 4 (right), Table 1). The first and second criteria provide a way to assess how well the compared approaches approximate the ground truth Mexican hat wavelet functions, while the third criterion measures the computational efficiency of the approaches. We compute the ground truth Mexican hat wavelet family in Eq. (5) with the complete Laplacian spectrum, and the intermediate diffusion times $t_n^{GT} = \log(nt)$, where $t$ is introduced in Algorithm 1. The evaluations are performed on all 100 shapes of the FAUST data set (remeshed to shapes with 5K vertices), using the average error at 10 sample locations, picked using farthest point sampling with random initialization.

As shown in Figures 4 (as well as Fig. 3 and Table 1 and Fig. 14 of the Appendix), our approach produces the most accurate approximation with respect to the ground truth and moreover is more computationally efficient than the competitors. Note that our approxi-mation of the ground truth diffusion wavelets is the best at small scales in terms of both $L_1$ and $L_\infty$ norms. This is especially important for partial matching where local details must be captured faithfully. Nevertheless, we note that [HQ12] provides the best representation of Mexican hat wavelets at large scales ($\geq 54$ scales).

### 3.3. Reconstruction and point-to-point map recovery

We use two approaches to perform point-wise signal recovery on shapes $\mathcal{N}$ and $\mathcal{M}$, equipped with their dictionaries $\Psi_\mathcal{S}^\mathcal{N}$ and $\Psi_\mathcal{S}^\mathcal{M}$. The choice of the approach is conditioned by the task we deal with.

**δ-function reconstruction on a single shape** In the case of δ-*function reconstruction* on a shape $\mathcal{M}$, the mapping $T$ that associates to each vertex index of $\mathcal{M}$ its image provided by $\Psi_\mathcal{S}^\mathcal{M}$ can be computed as

$$T = \underset{rows}{\arg\max} \Psi_\mathcal{S}^\mathcal{M} (\Psi_\mathcal{S}^\mathcal{M})^\dagger. \qquad (6)$$

The left pseudo-inverse of $\Psi_\mathcal{S}^\mathcal{M}$, denoted $(\Psi_\mathcal{S}^\mathcal{M})^\dagger$, is used here as the representation in dictionary space of all δ-functions of $\mathcal{M}$: the $k$-th column of $\Psi_\mathcal{S}^{\mathcal{M}\dagger}$ corresponds to the coordinates in the dictionary space of the δ-function located at vertex $k$. To convert back this dictionary representation in the basis of δ-functions on $\mathcal{M}$, $\Psi_\mathcal{S}^\mathcal{M}(\Psi_\mathcal{S}^\mathcal{M})^\dagger \in \mathbb{R}^{n_\mathcal{M} \times n_\mathcal{M}}$ is computed. The $k$-th column of the resulting $n_\mathcal{M} \times n_\mathcal{M}$ matrix is the image of the δ-function at vertex $k$. Taking the argmax over the rows of this matrix provides the location of the δ-function according to the dictionary $\Psi_\mathcal{S}^\mathcal{M}$. Since $\Psi_\mathcal{S}^\mathcal{M}$ is rank deficient, the computation of its pseudo-inverse via Eq. (6) is unstable. To remedy this, we use a Tikhonov (or ridge) regularization-like approach, that introduces the variable $\alpha$ in $T = \arg\max_{\text{rows}} \Psi_{\mathcal{S}}^{\mathcal{M}} \alpha$, which is the solution to $\arg\min_{\alpha} ||\Psi_{\mathcal{S}}^{\mathcal{M}} \alpha - I_{\mathcal{M}}||^2 + ||\Gamma\alpha||^2$, where $\Gamma$ is an $n_{\mathcal{M}} \times (|\mathcal{S}| \times n_{scales})$ matrix, whose columns contain the values $\frac{1}{k^2}$, with $k \in [1; n_{scales}]$ repeated $|\mathcal{S}|$ times each, and $I_{\mathcal{M}}$ is the identity matrix on $\mathcal{M}$. Then, the solution satisfies the linear system $\left[\Psi_{\mathcal{S}}^{\mathcal{M}}; \Gamma\right] \alpha = \left[I_{\mathcal{M}}; 0_{n_{\mathcal{M}} \times (|\mathcal{S}| \times n_{scales})}\right]$, where the semi-colon notation represents the column-wise concatenation of two matrices.

**δ-function transfer and shape matching for multiple shapes**
For *shape matching* (or *δ-function transfer*) from a source shape $\mathcal{M}$ to a target shape $\mathcal{N}$, we do not rely on the "spectrum" of the dictionaries. Instead, we use as a representation of a vertex $k$ on a shape $\mathcal{M}$ the *set of values* taken by each function constituting $\Psi_{\mathcal{S}}^{\mathcal{N}}$. In other words, instead of using $(\Psi_{\mathcal{S}}^{\mathcal{M}})^{\dagger}$ as a representation in our dictionary, we simply use $(\Psi_{\mathcal{S}}^{\mathcal{M}})^{\top}$. In this last representation, the embedding of the $k$-th vertex consists of the **value** taken by each wavelet of the dictionary at the $k$-vertex. Note that we *do not assume* the dictionary to be an orthogonal family (which it is not in most cases). To recover the mapping $T$ that associates to each vertex index of $\mathcal{M}$ its image on $\mathcal{N}$, we perform a nearest neighbor search $T = \text{NN-search}_{\text{rows}}\left(\Psi_{\mathcal{S}}^{\mathcal{N}}, \Psi_{\mathcal{S}}^{\mathcal{M}}\right)$, i.e., compute for each row of $\Psi_{\mathcal{S}}^{\mathcal{M}}$ its nearest neighbor among the rows of $\Psi_{\mathcal{S}}^{\mathcal{N}}$.

### 3.4. Theoretical guarantees

Our construction of the Mexican hat wavelets above inherits many attractive properties of the heat kernel, including isometry-invariance (due to invariance of the LBO), locality and its multi-scale nature. Moreover, as we demonstrate below, generically the relation to a single seed point through the Mexican hat wavelet $\psi_t^{\mathcal{M}}(p, x)$ is enough both to encode each point on the surface and to recover an isometry across a pair of shapes. Specifically, we call a point $p$ generic if it does not belong to any nodal set of the Laplace-Beltrami eigenfunctions, i.e. if $\phi_i(p) \neq 0$ for all $i$. As shown in [OMMG10], the set of generic points has full measure. Moreover, a surface is called generic if its Laplace-Beltrami eigenvalues are non-repeating. It is well-known [BU83] that an infinitesimal perturbation to a metric of any surface makes it generic. With these definitions, the following theorem guarantees that the uniqueness properties of the heat kernel also apply to our wavelet family construction.

**Theorem 1** Let $\mathcal{M}$ be a generic connected compact manifold without boundary and $p$ a generic point on $\mathcal{M}$. For any two points $x, y$, $x = y$ if and only if $\psi_t^{\mathcal{M}}(p, x) = \psi_t^{\mathcal{M}}(p, y)$ for all $t$. If $\mathcal{M}$ and $\mathcal{N}$ are two generic connected compact manifolds and $p$ a generic point on $\mathcal{M}$, then a map $T: \mathcal{M} \to \mathcal{N}$ where $T(p)$ is generic is an isometry if and only if $\psi_t^{\mathcal{M}}(p, x) = \psi_t^{\mathcal{N}}(T(p), T(x))$ for all $t$.

The proof of Theorem 1 follows the same reasoning as the proof of the main theorem in [OMMG10]. For the sake of completeness, we provide it in the Appendix (Section 8).

Theorem 1 implies that generically every point $x$ on a surface can be uniquely characterized by its relation to some fixed point $p$ via $\psi_t(p, x)$. Furthermore, an isometry can be recovered given a correspondence between a single pair of seed points, analogously to the heat kernel [OMMG10]. As we demonstrate below, however, our wavelet-inspired construction provides a more informative characterization in practice, while retaining the locality and multi-scale nature of the heat kernel.

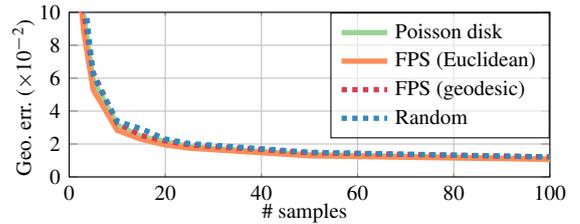

**Figure 5:** *Mean geodesic matching error (geo. err.) on the complete set of shapes from the SHREC'16 Partial cuts data set using various sampling strategies using* 25 *scales.*

## 4. Experimental analysis

We now study different aspects of the proposed approach, such as the sample placement, the number of scales, the computational robustness, and the choice of $t_{max}$. The details of all the datasets are provided in Section 8.2 of the Appendix.

**Sample selection** First, we assess the effect of the sampling strategy on the outcome of our method. To this end, we compare farthest point sampling (Euclidean and geodesic), Poisson disk sampling (computed using the gptoolbox Matlab package [J*18]) and random sampling on the SHREC'16 Partial cuts data set, with the geodesic matching error. According to Fig. 5, all sampling strategies behave in a similar fashion, and adding more samples improves the reconstruction error significantly by injecting more local information to the wavelet family. In all other experiments, we therefore use the Euclidean farthest point sampling strategy for its simplicity and more uniform localization of samples compared to random sampling. In the Appendix (Fig. 15, Sect. 8), we provide a complementary experiment on the SHREC'19 connectivity track data set with the δ-function reconstruction error, from which we draw identical conclusions.

**Sample robustness** Second, we verify the resilience of our approach to noise in the sample placement. We consider a set of 10 samples, among which we displace 1, 2, 3, 5 or all samples within a geodesic radius around the original sample location (noise radius). Six different scales are compared: 1, 2, 3, 5, 25 and 50 with noise radii varying between $1.0 \times 10^{-2}$ and $1.0 \times 10^{-1}$ of the greatest geodesic distance on the shape. Fig. 6 summarizes our results collected on the complete SHREC'16 Partial cuts data set. As a baseline, we display the matching error produced by using a dictionary of wavelets from [HQ12] and heat kernel functions, both with 25 scales. This experiment furthermore illustrates the representative power of our approach in the case of partial shape matching.

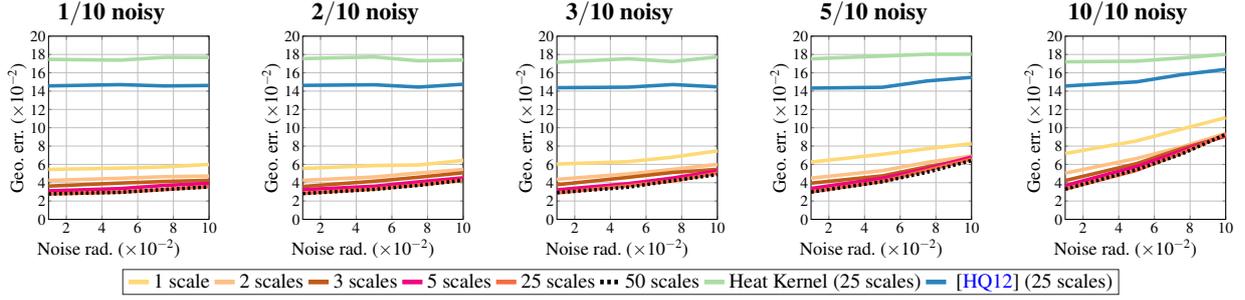

**Figure 6:** *Geodesic matching error as a function of the noise level applied to the positioning of the samples on the complete SHREC'16 Partial cuts data set. The level of noise is given as a noise radius (noise rad.), representing the geodesic disc centered around the original sample, which the noisy sample is drawn from. The geodesic radius is expressed as a fraction of the maximum geodesic distance. In each column, an increasing number of samples are noisy (from left to right: 1, 2, 3, 5 and 10, out of a total of 10 samples).*

**Table 2:** *Mean geodesic error on 200 shape pairs of the FAUST data set and 212 pairs of the TOSCA Isometric data set (original, with edges flipped and remeshed to 5K vertices in both cases), using 25 scales and 10 samples.*

| Data set | Mean geodesic error |
|---|---|
| Faust (original) | $9.96 \times 10^{-2}$ |
| Faust (remeshed) | $\mathbf{6.15 \times 10^{-2}}$ |
| Faust (edges flipped) | $9.77 \times 10^{-2}$ |
| TOSCA (original) | $\mathbf{4.78 \times 10^{-2}}$ |
| TOSCA (remeshed) | $4.99 \times 10^{-2}$ |
| TOSCA (edges flipped) | $6.02 \times 10^{-2}$ |

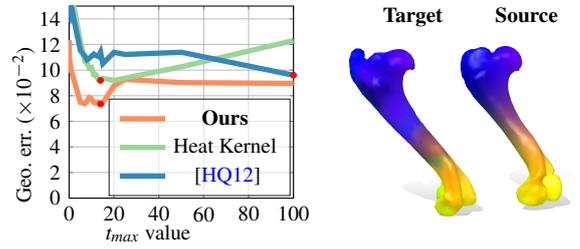

**Figure 7:** *Left: mean geodesic error on 30 shape pairs of the humerus bones data set, remeshed to 1k shapes. The minimum matching error of $7.35 \times 10^{-2}$ is attained for $t_{max} = 14$ (red dot). Right: illustration of a matching estimated between two bones using the best $t_{max}$ value. Corresponding points are depicted with the same color. The target shape was rescaled by a factor of $\times 0.8$ to match the size of the source shape.*

**Number of scales** Fig. 6 empirically indicates that choosing 25 scales is a good trade-off between robustness to noise and computational efficiency, especially when the sample position is inaccurate.

**Robustness to topological changes** We verify the robustness of our method to topological changes by comparing three versions of the FAUST and TOSCA Isometric data sets: (i) the original data sets, with shapes counting respectively 6890 and around 25K vertices, (ii) the data sets remeshed to shapes with close to 5K vertices and (iii) the original data sets with random edge flips applied to 12.5% of the original edges. Table 2 demonstrates that our computation is robust to these changes leading to similar low error in all these scenarios.

**Choice of $t_{max}$** The maximum diffusion time $t_{max}$ remains a free parameter of our method. Throughout the experiments that we present, we choose to fix its value to 1, since it provides good results on the data sets that we are using. However, selecting its value depending on the shape could allow to improve the quality of the matching, in particular in situations where the samples cannot be placed regularly on the shape's surface. To illustrate this aspect, we conducted an experiment on the humerus bones data set. All shapes have been remeshed to count 1K vertices. Fig. 7 shows that the geodesic error varies substantially depending on $t_{max}$. Its value is the smallest for $t_{max} = 14$. Furthermore, to highlight the representative power of our construction, we use the heat kernel and the diffusion wavelets of [HQ12] as a baseline. For all diffusion time selected in this experiment, we outperform both approaches.

## 5. Experimental Comparisons

To illustrate the benefits of the proposed approach, we discuss self- (Sect. 5.1) and regular (Sect. 5.2) shape matching, and compare our performance to the heat kernel (Sect. 5.3).

### 5.1. Self-matching

One feature of our approach is that it provides a better representation for δ-functions. With only a small number of sample points, we provide an approximation of δ-functions that is significantly more accurate than traditional functional bases, such as the LBO eigenfunctions. To illustrate this aspect, we consider *self-matching*, in which we evaluate the expressive power of our family in reconstructing δ functions, thereby matching the vertices of a shape to itself. Fig. 8 presents the results obtained on all shapes of the SHREC'16 Partial cuts data set and all shapes of the TOSCA non Isometric data set (remeshed version). The evaluation indicates the error in terms of geodesic radius, identically to the procedure in [KLF11] but using the same shape as source and target.

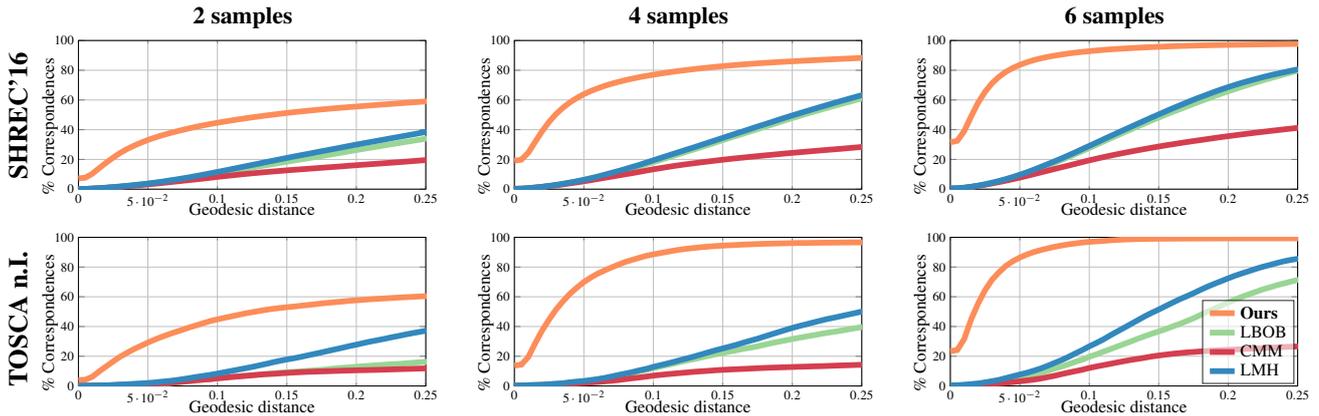

**Figure 8:** *Geodesic error (self-matching) on all shapes of the SHREC'16 Partial cuts data set (top row) and on all shapes of the TOSCA non-isometric data set (category 8) (bottom row). Averaged values are reported respectively in Tables 3 and 4. Each column is with a different number of sample points/non constant basis functions. Our functional dictionary recovers points on the surfaces much more accurately for the same basis budget.*

**Table 3:** *Average geodesic error for the SHREC'16 Partial cuts data set (self-matching), corresponding to the top row of Fig. 8.*

| # Gt Corres. | 2 | 4 | 6 |
| --- | --- | --- | --- |
| LBOB | $3.84 \times 10^{-1}$ | $2.27 \times 10^{-1}$ | $1.69 \times 10^{-1}$ |
| LMH | $4.09 \times 10^{-1}$ | $2.32 \times 10^{-1}$ | $1.80 \times 10^{-1}$ |
| CMM | $7.03 \times 10^{-1}$ | $6.22 \times 10^{-1}$ | $5.08 \times 10^{-1}$ |
| **Ours** | $\mathbf{2.28 \times 10^{-1}}$ | $\mathbf{1.00 \times 10^{-1}}$ | $\mathbf{3.75 \times 10^{-2}}$ |

**Table 4:** *Average geodesic error (self-matching) for all 24 shapes from the TOSCA non isometric data set (category 8), corresponding to the bottom row of Fig. 8.*

| # Gt Corres. | 2 | 4 | 6 |
| --- | --- | --- | --- |
| LBOB | $5.55 \times 10^{-1}$ | $3.17 \times 10^{-1}$ | $2.03 \times 10^{-1}$ |
| LMH | $4.73 \times 10^{-1}$ | $2.98 \times 10^{-1}$ | $1.81 \times 10^{-1}$ |
| CMM | $7.66 \times 10^{-1}$ | $7.34 \times 10^{-1}$ | $5.80 \times 10^{-1}$ |
| **Ours** | $\mathbf{3.94 \times 10^{-1}}$ | $\mathbf{6.9 \times 10^{-2}}$ | $\mathbf{3.1 \times 10^{-2}}$ |

To build our family of functions, we use 2, 4, or 6 samples, placed using Euclidean farthest point sampling, 25 scales, $t_{max} = 1$ for the maximum diffusion scale, and the point-to-point map conversion for δ-function reconstruction described in Sect. 3.3. We compare the performance of our dictionary to the LBO eigenfunctions basis (LBOB), the Localized Manifold Harmonic basis (LMH) [MRCB18] and the Compressed Manifold Modes (CMM) [NVT*14], using $|\mathcal{S}| + 1$ basis functions, to take into account the constant function of the LBOB. For each of these methods, the δ-function location is determined by taking the position at which its approximation in the basis considered is maximal. According to the mean geodesic error for all approaches (Tables 3, 4), the proposed method outperforms significantly the other methods.

### 5.2. Pairwise shape matching

In a more practical application, we study how well our family of functions recovers δ-functions basis after being transferred from a source $\mathcal{M}$ to a target shape $\mathcal{N}$ via shape matching. This corresponds to the scenario of extending a set of known seed point correspondences to the entire shapes.

For our family of functions, we employ the same setup as for the δ-function reconstruction with a few adjustments: we use 3, 10 or 20 samples and the transfer point-to-point conversion introduced in Sect. 3.3. The remaining parameters stay identical (25 scales and $t_{max} = 1$).

In all cases we assume that the ground truth correspondences between the bases. For landmark-aware bases such as ours, this means that we assume the knowledge of ground truth correspondences between $|\mathcal{S}|$ sample points on source and target shapes. For global bases such as the LBO, we assume the ground truth correspondence between $|\mathcal{S}|$ first non-constant basis functions. In the latter setting, we follow the procedure used in [MRCB18] and leverage this known correspondence to build a ground truth functional map [OBCS*12], given as $C^{gt} = \phi_{\mathcal{N}}^{\top} A_{\mathcal{N}} \Pi^{gt} \phi_{\mathcal{M}}$, where $\Pi^{gt}$ is the ground truth point-to-point map. We then use this ground truth functional map $C^{gt}$ to compute the dense point-to-point map following the the standard nearest-neighbor procedure [OCB*17] (Chapter 2).

Fig. 9 presents the results on pair of shapes from the same data sets as for the δ-function reconstruction experiment and comparing again against the LBOB, LMH and CMM bases. The evaluation, performed according to the standard protocol proposed in [KLF11], indicates the error in terms of geodesic radius. According to the average values in Tables 5, 6, our dictionary outperforms the competing bases by a substantial margin.

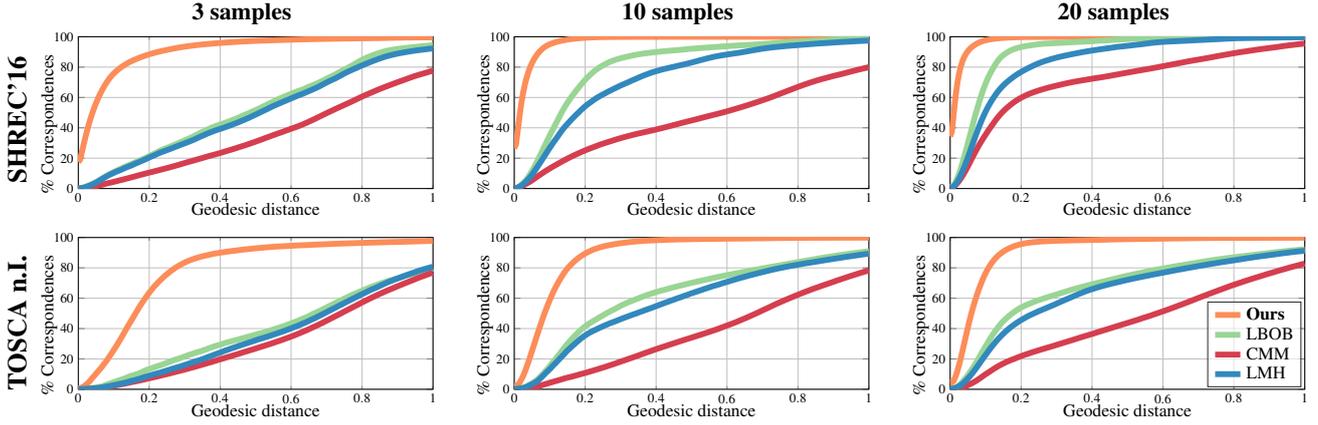

**Figure 9:** *Matching geodesic error on all pairs of the SHREC'16 Partial cuts data set ("SHREC'16", top row, see averaged values in Table 5) and 190 shape pairs of the TOSCA non-isometric data set ("TOSCA n.I.", bottom row, see averaged values in Table 6) using an increasing number of ground truth sample points/non constant basis functions correspondences. On all plots, the x-axis represents the normalized geodesic distance and the y-axis is the fraction of correspondences in percent.*

**Table 5:** *Average geodesic error (partial shape matching) for all shape pairs of the SHREC'16 Partial cuts data set, corresponding to the bottom row of Fig. 9.*

| # Gt Corres. | 3 | 10 | 20 |
|---|---|---|---|
| LBOB | $4.94 \times 10^{-1}$ | $1.93 \times 10^{-1}$ | $9.68 \times 10^{-2}$ |
| LMH | $5.22 \times 10^{-1}$ | $2.75 \times 10^{-1}$ | $1.59 \times 10^{-1}$ |
| CMM | $7.06 \times 10^{-1}$ | $6.01 \times 10^{-1}$ | $3.02 \times 10^{-1}$ |
| **Ours** | $\mathbf{8.88 \times 10^{-2}}$ | $\mathbf{2.82 \times 10^{-2}}$ | $\mathbf{1.94 \times 10^{-2}}$ |

**Table 6:** *Average geodesic error (full shape matching) for 190 shape pairs from the TOSCA non-isometric data set, corresponding to the bottom row of Fig. 9.*

| # Gt Corres. | 3 | 10 | 20 |
|---|---|---|---|
| LBOB | $6.70 \times 10^{-1}$ | $4.04 \times 10^{-1}$ | $3.44 \times 10^{-1}$ |
| LMH | $6.94 \times 10^{-1}$ | $4.51 \times 10^{-1}$ | $3.82 \times 10^{-1}$ |
| CMM | $7.36 \times 10^{-1}$ | $6.87 \times 10^{-1}$ | $5.98 \times 10^{-1}$ |
| **Ours** | $\mathbf{2.14 \times 10^{-1}}$ | $\mathbf{1.08 \times 10^{-1}}$ | $\mathbf{7.97 \times 10^{-2}}$ |

### 5.3. Comparison with the heat kernel

The construction of our functions is closely related to those provided by the heat kernel. While both function types share the ability to characterize uniquely every point on a surface, our heat kernel derivatives are more informative in practice. To assess this practical advantage, we conduct the following experiment on a set of 10 pairs of the dog class from the TOSCA data set. Given an increasing number of samples, we compute for each pair of shapes the AUC (Area Under the Curve: the probability that a point is matched with an error less or equal to 0.25 in normalized geodesic distance) and the mean geodesic error using the proposed family and the heat kernel, associated with the conversion of a point-to-point map.

We setup our dictionary using the same parameters as for the δ-function transfer experiment, using ground-truth correspondences between the sample points on the source and target shapes. The quantitative and qualitative evaluation of this experiment is depicted in Fig. 10. Relying on a diffusion process to define both families of functions, we emphasize that this experiment can be seen as an additional comparison to standard diffusion wavelets. This result highlights that heat kernel derivatives are more informative than heat kernel functions.

### 6. Application to Partial Shape Matching

As the main application of our method, we tackle the problem of partial shape matching, one of the challenging scenarios in non-rigid shape matching. We experiment on the SHREC'16 Partial Correspondence benchmark [CRB*16] (Sect. 6.2) and on a new set of partial shapes, namely FARM partial (Sect. 6.1). If not specified, we adopt as sparse set of correspondences for our approach the fully automatic result obtained with the pipeline proposed in [RBA*12], initialized with the SHOT descriptor [TSDS10]. We highlight that our main contribution is an informative, localized functional family, which leads to a remarkably simple and effective shape matching approach. In the following experiments we compare our approach to existing full-fledged optimization and learning-based strategies, specifically designed for partial matching. Thus, the simplicity and efficiency of our approach should be taken into account when comparing to more advanced and highly tuned methods.

### 6.1. FARM partial dataset

We first evaluate our method on the FARM partial dataset. This dataset contains partiality and shapes undergoing non-isometric deformations and extremely different connectivities. This makes this dataset particularly challenging as many shape matching pipelines are known to overfit to similar mesh connectivities. On the left of Fig. 11, we show a quantitative comparison on FARM partial to state-of-the-art Partial Functional Maps [RCB*17] (PFM) method. For a fair comparison, we additionally evaluate the performance of PFM when it is initialized with the same sparse correspondence that we exploit to generate our family of functions (PFM sparse). We also provide a qualitative illustration of the computed maps in Figure 12.

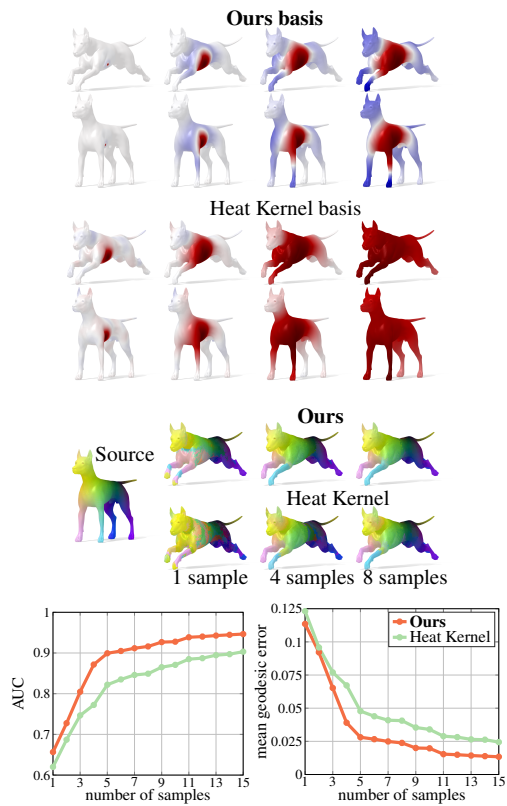

**Figure 10:** *(1st, 2nd Rows) Comparison with heat kernels in point-to-point map conversion with the same number of scales and a small number of point samples. Four scales of the heat diffusion from a sample on a pair of shapes; the colormap ranges from blue (negative) to red (positive) with values close to zero in white. (3rd Row) Qualitative comparison of the resulting maps for 1, 4, 8 samples (left to right), using color correspondence to show the resulting point-to-point map between a source and a target shape. (4th Row) Performance of our approach compared to the heat kernel in terms of Area Under the Curve (AUC) and mean geodesic error. Results are averaged over 10 pairs of the dog class from the TOSCA data set. The evaluation highlights the better performance of our representation over the heat kernel.*

Note that the state-of-the-art PFM does not perform well on these challenging pairs. In contrast, our method is robust, significantly simpler and more efficient, leading to a dramatic improvement in accuracy.

### 6.2. SHREC'16 Partial Cuts benchmark

We also evaluate our method In the evaluation on the SHREC'16 Partial Cuts data set, where each partial shape is matched to the full shape of the same shape category. Remark that this dataset contains shape pairs undergoing near-isometric deformations, which are well-captured by the LBO basis.

The quantitative evaluation is shown in Fig. 11 (middle and right). On the left, we compare our approach on the entire cuts set from SHREC'16 [CRB*16] with all the methods that were considered in the challenge. Our performance is comparable with partial functional maps [RCB*17] (PFM) the state-of-the-art for partial matching. The constrained optimization performed by PFM produces more accurate correspondences because it is able to solve the inaccuracies contained in the initial sparse correspondence. Note, however, that due to the way the data set was produced, the shape pairs of this data set have similar connectivity, which is a known factor of overfitting for shape matching techniques.

In Fig. 11 (right), we compare our approach to PFM when both are initialized with 20 and 30 ground-truth correspondences only on the cat class. As can be seen, if the sparse correspondences are correct, our method is comparable to PFM and even better. We highlight that this is the only evaluation in which we use a ground-truth initialization. According to the qualitative results in Fig. 13, our performance is comparable to PFM.

**Computational Efficiency** When considering the computational efficiency (in seconds), our method outperforms PFM by a significant margin. On the complete SHREC16 data set, PFM sparse takes on average **138.2s** per shape pair, PFM takes **240.9s**, while our method requires **46.2s**. Moreover, the sparse set of samples takes on average 38.7s per shape to be computed. Therefore, most of the computation overhead lies in this preprocessing step for our method. Once a set of sparse correspondences is available, we require an average computation time of **7.5s** per shape pair, which represents an improvement of 13x compared to PFM sparse (**99.5s**).

### 7. Conclusions

In this work, we have proposed an extension to the basic diffusion (heat kernel) construction by considering its derivatives in time, or equivalently in space. The resulting family of diffusion-based Mexican hat wavelets is local and allows to find accurate point-to-point correspondences between shapes; this includes particularly challenging settings such as partial shape correspondence, and matching between shapes with highly different triangulations. At the same time, the efficient use of diffusion-based methods allows to solve these difficult problems at a fraction of the computational cost compared to other approaches. We further proved that our functions inherit properties of the heat kernel map, such as the ability of only requiring one sample point to recover an isometry.

Our experiments on δ-function reconstruction and transfer indicate that our family can be thought of as an over-complete basis that provides a richer functional representation power compared to LBO eigenfunctions, diffusion functions, and other bases. Moreover, the application of wavelet-like functions on partial and large-scale shapes show promising results compared to state-of-the-art methods, especially when taking into consideration its simplicity.

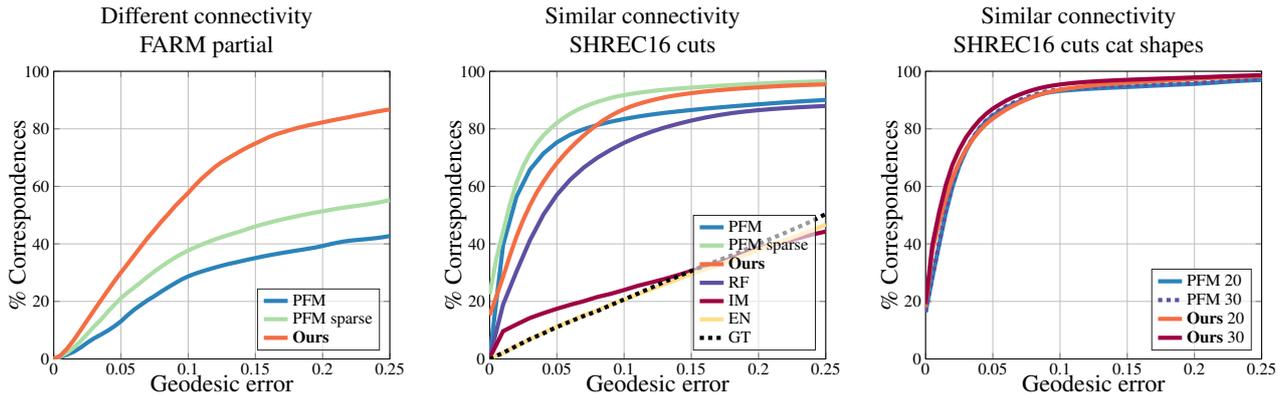

**Figure 11:** *Quantitative comparison on the FARM partial data set (shapes with different connectivity) and SHREC'16 partial cut benchmark (composed of shapes with similar connectivity). In all plots, the x-axis is the mean geodesic distance to the ground truth. Abbreviations used: PFM (partial functional maps), PFM sparse (PFM initialized with the same sparse correspondence used to compute our frame), RF, IM, EN, GT. For PFM and ours applied to SHREC'16 cuts on the cat shape, an additional number specifies the number of ground-truth correspondences that were used for initialization (20 or 30).*

The *main limitation* of our approach currently lies in the dependency on an initial sparse correspondence, which is assumed to be roughly accurate. Although further progress in deformable *sparse* matching would have a direct and positive impact on our method, we believe that this problem can be solved jointly within our frame calculation algorithm, and leave this challenge as an exciting direction of future research.

**Acknowledgments** We thank the Reviewers for their comments, which helped us to improve the technical and experimental parts of the article. Parts of this work were supported by the KAUST OSR Award No. CRG-2017-3426, the ERC Starting Grant No. 758800 (EXPROTEA), and the ERC Starting Grant No. 802554 (SPEC-GEO).

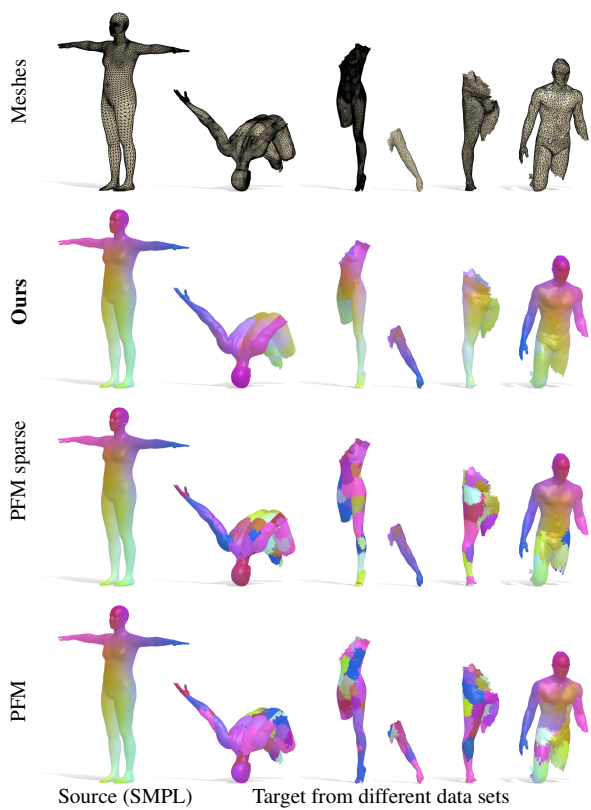
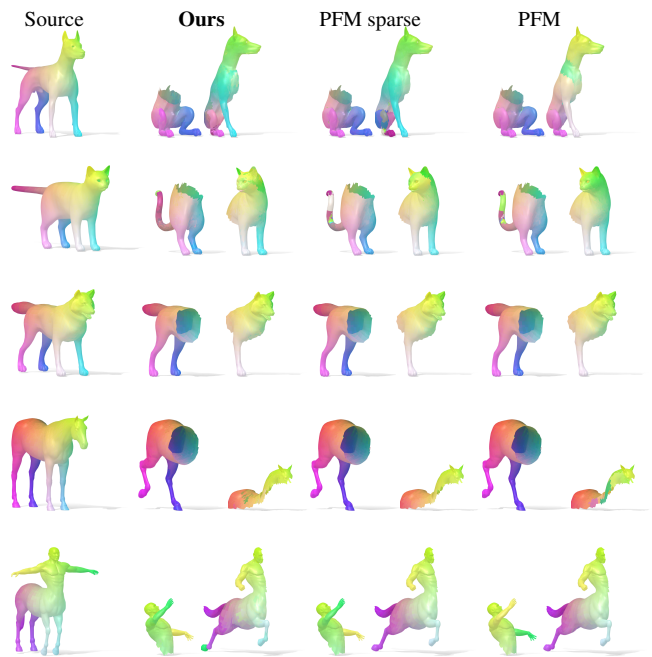

**Figure 12:** *(First row) Different mesh connectivity. (Second-fourth rows) Qualitative comparison on the FARM partial data set between our approach and the PFM in is original version (PFM) and initialized with the sparse correspondence that we adopt for the definition of our family of functions (PFM sparse).*

**Figure 13:** *Qualitative comparison on the SHREC'16 partial cut benchmark for 5 classes (wolf, horse, centaur, dog, cat) between our approach, the PFM (original version), and the PFM initialized with the sparse correspondence that we adopt for the definition of our family of functions (PFM sparse). The resulting point-to-point mapping is displayed through color correspondence. Our approach, despite its simplicity, is comparable to PFM.*

## 8. Appendix

### 8.1. Proof of Theorem 1

Both statements of the theorem follow directly from the spectral expansion of $\psi_t^M(p,x) = \sum_i \lambda_i \exp(-t\lambda_i)\Phi_i(p)\Phi_i(x)$ and the following lemma, proved in [OMMG10] (Lemma 3.2, Remark 3.3).

**Lemma 1** Given two strictly increasing sequences $\lambda_i$ and $\mu_i$ of non-negative numbers that tend to infinity, if $a(t) = \sum_i a_i \exp(-t\lambda_i)$ and $b(t) = \sum_i b_i \exp(-t\mu_i)$ where $a_i, b_i \neq 0$ then $a(t) = b(t)$ for all $t$ if and only if $\lambda_i = \mu_i$ and $a_i = b_i$ for all $i$.

Applying this lemma to the spectral expansion of $\psi$ while recalling that only the first eigenvalue on a connected manifold is zero, we immediately get the first statement of Theorem 1 (using the same proof as of Theorem 3.1 in [OMMG10]). The second statement follows from the same argument as Theorem 3.5 in [OMMG10]). Namely, by first applying this lemma to $x = p$, and $y = q$, which implies preservation of eigenvalues and second to other points on the surface implying preservation of all but first eigenfunction. Together this implies that $T$ preserves $\psi$ if and only if it preserves the Laplace-Beltrami operator, which is equivalent to an isometry.

### 8.2. Data sets

The *FAUST training* data set [BRLB14] consists of 100 human shapes, with 10 different humans in 10 different poses. All shapes have a consistent manifold mesh structure, with 6890 vertices.

The *TOSCA* data set [BBK08] is composed of 80 shapes of various categories: 11 cats, 9 dogs, 3 wolves, 8 horses, 6 centaurs, 4 gorillas, 12 female figures, and 2 different male figures, in 7 and 20 poses. The mean vertex count is about 50K. If not explicitly mentioned, the shapes of this data set are remeshed to count around 5K vertices each. In the *TOSCA Isometric* data set, we consider shape pairs within the same category (e.g., cats matched to cats), whereas in the *TOSCA non-Isometric* data set, we consider matches between the gorilla shapes and the two human categories (male and female).

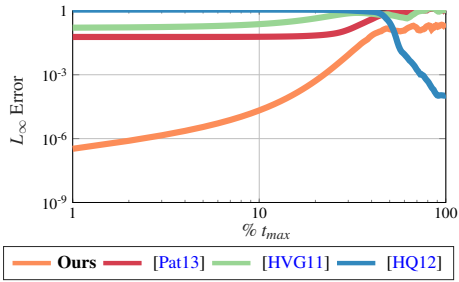

**Figure 14:** *Comparison of the $L_\infty$ error to the ground-truth diffusion wavelets for various Mexican hat wavelets. See Table 1.*

The *Humerus Bones* data set is composed of a collection of 15 humerus bone models of wild boars acquired using a 3D sensor. Each bone was scanned independently, with 24 consistent landmarks provided by experts in the field [GM13] on each shape. The original resolution of the shapes is around 25K vertices.

The *SHREC'16 partial cuts correspondence benchmark* [CRB*16], is the most adopted data set for non-rigid shape matching. The shapes belong to 8 different classes (5 animals and 3 humans). Each class contains pose deformations and partiality transformations, i.e., regular cuts and irregular holes. We limit our evaluation to the cuts set (120 pairs), which are resampled independently to ∼10K vertices and share a similar density.

The *SHREC'19 connectivity track benchmark* [MMR*19] is composed of 44 shapes of humans. The complete benchmark consists in 430 shape pairs composed by meshes that represent deformable human body shapes. Shapes belonging to these categories undergo changes in pose and identity. The meshes exhibit variations of two different types: density (from 5K to 50K vertices) and distribution (uniform, non-uniform). For each shape, the full SMPL model [LMR*15] (6890 vertices) serves as our ground-truth.

The *FARM partial data set* is a collection of partial shapes that we extract from a subset of 5 meshes of the SHREC'19 connectivity track [MMR*19]. These shapes belong to different data set: TOSCA [BBK08] (around 50K vertices), SPRING [YYZ*14] (12.5K vertices) and K3DHUB [XZC18] (around 10K vertices). We randomly cut five patches from each of these shapes (Fig. 12) and each partial mesh is matched with the full SMPL model [LMR*15] (6890 vertices). The ground-truth correspondence is extended to these partial shapes from the FARM registration [MMRC18], which provides a ground truth dense correspondence between SMPL and each of the full shapes involved.

### 8.3. Comparison to other wavelets & sampling

**Scalability** As a complement to Fig. 3, Table 7 displays the computation time required by [HQ12] and our approach for various order of magnitude. In all cases, the proposed method outperforms [HQ12] by at least a factor 2.

$L_\infty$ **error** In addition to the $L_2$ error displayed in Fig. 4, we also measured the $L_\infty$ of the error, using the same experimental setup

**Table 7:** *Computation time (in sec.) of [HQ12] compared to our approach on 5 shapes (Fig. 3) from the SHREC'19 data set.*

| # vertices | [HQ12] (s.) | **Ours** (s.) | Improv. |
|---|---|---|---|
| $10^3$ | 2.03 | **0.38** | ×5.34 |
| $10^4$ | 9.00 | **3.05** | ×2.95 |
| $10^5$ | 73.39 | **25.00** | ×2.94 |
| $10^6$ | 524.63 | **247.02** | ×2.12 |

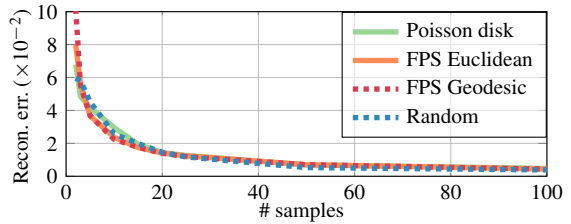

**Figure 15:** *Mean reconstruction error (recon. err.) as a function of the number of samples (# samples) on all δ-functions for 26 shapes of the SHREC'19 connectivity track data set, using 25 scales.*

as in Sect. 3.2. The result, shown in Fig. 14, is similar to what we observe for the $L_2$ error.

**Sampling strategy** To furthermore illustrate the independence of our method to sampling, we compute the mean reconstruction error as a function of the number of samples on all δ-functions for 26 shapes of the SHREC'19 connectivity track. The outcome of this experiment (Fig. 15) is similar to what we observed in Sect. 4.